\documentstyle[11pt,twoside,epsfig]{article}
\def\Journal#1#2#3#4{{#1} {\bf #2}, #3 (#4)}

\def\NPB{{\em Nucl. Phys.} B}
\def\PLB{{\em Phys. Lett.}  B}
\def\PRL{\em Phys. Rev. Lett.}

\def\als{\alpha_s}

\def\be{\begin{equation}}
\def\ee{\end{equation}}
\def\bea{\begin{eqnarray}}
\def\eea{\end{eqnarray}}

\def\epp{\varepsilon^\prime/\varepsilon}
\def\nn{\nonumber}
\newcommand{\klpn}{K_L \to \pi^0 \nu \bar \nu}
\newcommand{\kppn}{K^+ \to \pi^+ \nu \bar \nu}
\newcommand{\kmm}{K_L \to \mu^+ \mu^-}
\newcommand{\kpe}{K_L \to \pi^0 e^+ e^-}
\def\Im{\mathop{\mbox{Im}}}
\def\Re{\mathop{\mbox{Re}}}

\newcommand{\mtb}{\overline{m}_{t}}
 
\def\r#1{(\ref{#1})}
\begin{document}
\thispagestyle{empty}
\begin{flushright}
 TUM-HEP-350/99 \\
 June 1999
\end{flushright}
\vskip1truecm 
\centerline{\Large\bf CONSTRAINTS ON AN ENHANCED }
\centerline{\Large\bf $\bar s d Z$  VERTEX FROM $\epp$\footnote[1]{\noindent
    Talk given at the XXXIVnd Rencontres de Moriond ``ELECTROWEAK
    INTERACTIONS AND UNIFIED THEORIES'', Les Arcs, France, 13-20 March
    1999.}}  
\vskip1truecm 
\centerline{\large\bf Luca Silvestrini} 
\bigskip 
\centerline{\sl Technische
  Universit\"at M\"unchen, Physik Department} 
\centerline{\sl D-85748
  Garching, Germany} 
\vskip1truecm 
\centerline{\bf Abstract} 
We analyze rare kaon decays in models in which
the dominant new effect is an enhanced $\bar s d Z$ vertex $Z_{ds}$.
We point out that in spite of large theoretical uncertainties the
CP-violating ratio $\epp$ provides at present the strongest
constraint on $\Im Z_{ds}$. Assuming $1.5 \cdot 10^{-3} \le \epp \le
3 \cdot 10^{-3}$ and Standard Model values for the CKM parameters we
obtain the bounds ${\rm BR}(\klpn)\le 3.9 \cdot 10^{-10}$ and ${\rm
  BR}(\kpe)_{\rm dir} \le 7.9 \cdot 10^{-11}$. Using the bound on
$\Re Z_{ds}$ from $\kmm$ we find ${\rm BR}(\kppn) \le 2.6 \cdot
10^{-10}$.  We also discuss new physics scenarios in which in
addition to an enhanced $\bar s d Z$ vertex also neutral meson
mixing receives important new contributions. In this case our most
conservative bounds are ${\rm BR}(\klpn)\le 1.8 \cdot 10^{-9}$,
${\rm BR}(\kpe)_{\rm dir} \le 2.8 \cdot 10^{-10}$ and ${\rm
  BR}(\kppn) \le 6.1 \cdot 10^{-10}$. The dependence of the upper
bound for ${\rm BR}(\klpn)$ on the CKM parameters is also analyzed.

\vfill 
\newpage 

\section{Introduction}

Flavour Changing Neutral Current (FCNC) and CP-violating processes
represent a very powerful tool to search for low-energy effects of
physics beyond the Standard Model (SM). In fact, the gauge structure
and particle content of the SM, together with the requirement of
renormalizability, imply the absence of tree-level FCNC and the strong
suppression of these processes at higher loops via the
Glashow-Iliopoulos-Maiani (GIM) mechanism. If now the SM is considered
as the low-energy limit of a more fundamental theory, the exchange of
new particles of mass $M_X$ will generate a series of effective
operators of higher dimension, suppressed by inverse powers of $M_X$.
These operators will in general violate the accidental symmetries of
the SM, and in particular they will introduce effective FCNC vertices.
The experimental results on FCNC processes will then in general imply
very stringent bounds on these effective vertices.

Rare Kaon decays are among the cleanest FCNC and CP violating
low-energy transitions, since they are essentially free of hadronic
uncertainties and directly probe the product $\lambda_t=V_{ts}^*
V_{td}$ of Cabibbo-Kobayashi-Maskawa (CKM) matrix elements, the
imaginary part of which is proportional to the invariant measure of CP
violation in the SM.

It is therefore interesting to consider possible new physics
contributions to rare $K$ decays. To discuss these effects in full
generality, let us analyze the structure of the possible effective
FCNC $\bar s d$ vertices generated by new physics contributions
arising at the scale $M_X$. In general, the following FCNC effective
couplings will arise via penguin diagrams: $\bar s d \gamma$,
$\bar s d g$ and $\bar s d Z$, while effective four-fermion operators
$\bar s d \bar f f$ will be generated via box diagrams. Naive power
counting implies the following structure for the above contributions:
\begin{eqnarray}
  {\rm Photon~penguins:} &&
  \frac{1}{M_X^2} \bar s_A \gamma^\mu d_A (q^2 g_{\mu \nu} - q_\mu q_\nu)
  A^\nu \to 
  \frac{1}{M_X^2} \bar s_A \gamma^\mu d_A \bar f \gamma^\mu f \nn \\
  {\rm Gluon~penguins:} &&
  \frac{1}{M_X^2} \bar s_A \gamma^\mu t_a d_A (q^2 g_{\mu \nu} - q_\mu
  q_\nu) A^{\nu}_a \to
  \frac{1}{M_X^2} \bar s_A \gamma^\mu d_A \bar f \gamma^\mu f \nn \\
  {\rm Box~diagrams:} && 
  \frac{1}{M_X^2} \bar s_A \gamma^\mu d_A \bar f_B \gamma^\mu f_B \nn \\
  Z~{\rm penguins:} &&
  \bar s_A \gamma^\mu d_A Z_\mu \to
  \frac{1}{M_Z^2} \bar s_A \gamma^\mu d_A \bar f_B \gamma^\mu f_B,
  \label{eq:naive}
\end{eqnarray}
where $A,B=L,R$. Due to electroweak symmetry breaking, a dimension
four FCNC $\bar s d Z$ coupling is allowed and one would naively
conclude from eq.~(\ref{eq:naive}) that there is no decoupling of new
physics contributions to the $\bar s d Z$ effective vertex, leading to
a four fermion operator suppressed only by $1/M_Z^2$, and not by
$1/M_X^2$ as for the other contributions in eq.~(\ref{eq:naive}).
However, the above conclusion is incorrect, since to generate a FCNC
$\bar s d Z$ coupling some electroweak breaking effect must be present
in the loop, thereby introducing a suppression factor of order
$M_Z^2/M_X^2$, which leads to a $1/M_X^2$ suppression of the
four-fermion operator. In other words, the last line of
eq.~(\ref{eq:naive}) should be replaced by
\begin{equation}
  \label{eq:naive2}
  Z~{\rm penguins:} \qquad
  \frac{M_Z^2}{M_X^2} \bar s_A \gamma^\mu d_A Z_\mu \to
  \frac{1}{M_X^2} \bar s_A \gamma^\mu d_A \bar f_B \gamma^\mu f_B.
\end{equation}

The above discussion is only based on a dimensional argument. However,
one has to remember that in order to generate these FCNC effective
operators, some flavour violation must be present in the diagrams. In
a large class of extensions of the SM, such as SUSY models for
example, a generalization of the GIM mechanism is present, leading to
an additional suppression factor of order $m_q^2/M_X^2$, where $m_q$
is a quark mass. Now, it is interesting to note that the same factor
$m_q^2/M_X^2$ can provide, in the case of the $Z$ penguin, both
electroweak and flavour breaking, leading to an enhancement of $Z$
contributions with respect to other penguin and box contributions.
In other words, in this kinds of models where a ``hard'' GIM
suppression is present, one obtains the following
hierarchy:
\begin{equation}
  Z~{\rm penguins:}~\frac{1}{M_Z^2}\frac{m_q^2}{M_X^2} \bar 
  s_A \gamma_\mu d_A \bar f_B \gamma^\mu f_B; \qquad
  {\rm other~penguins,~boxes:}~
  \frac{1}{M_X^2}\frac{m_q^2}{M_X^2} \bar 
  s_A \gamma_\mu d_A \bar f_B \gamma^\mu f_B,
  \label{eq:classi}
\end{equation}
i.e. a suppression of $M_Z^2/M_X^2$ of all the other contributions with
respect to the $Z$ one. 

This has been pointed out by Colangelo and Isidori in the framework of
general SUSY models, where the authors have shown that a particular
contribution can enhance the $\bar s d Z$ effective vertex by up to one
order of magnitude with respect to the SM, even for relatively large
masses of the superpartners, where all other effects are small~\cite{CI}.

Motivated by the above arguments, we consider in the following
a generic class of models in which an enhanced $\bar s d Z$ effective
coupling is present, and analyze the phenomenological constraints that
can be placed on this vertex and its consequences for rare $K$
decays. We follow closely the discussion in ref.~\cite{BS},
updating the results presented there in the light of the new
experimental data and of the analysis in ref.~\cite{epp}.

\section{The effective $\bar s d Z$ vertex and rare $K$ decays}
\label{sec:effvert}

We define the effective $\bar s d Z$ vertex as follows:
\begin{equation}
  \label{eq:Wds}
  {\cal L}^Z_{\rm FC} = \frac{G_F}{\sqrt{2}} \frac{e}{2 \pi^2} M_Z^2
  \frac{\cos \Theta_W}{\sin \Theta_W} Z_{ds} \bar s \gamma_\mu
  (1-\gamma_5) d~ Z^\mu\,+\, {\rm h.c.}
\end{equation}
where $Z_{ds}$ is a complex coupling. In the Standard Model one has
$Z_{ds}^{\rm SM} = \lambda_t C_0(x_t)$ with
$x_t={\mtb^2}/{M_W^2}$,
where $\lambda_t=V_{ts}^* V_{td}$ and $C_0(x_t)$ is a real function
which for the central value of the top quark mass, $\mtb(m_t)=166$
GeV, equals $0.79$. Its explicit expression can be found in
\cite{bratios}.

From the standard analysis of the unitarity triangle, we find,
neglecting the error in $\mtb$,
\begin{eqnarray}
  \label{eq:smlt}
  \Im \lambda_t = (1.34 \pm 0.30) \cdot 10^{-4}\,, &\qquad& \Re
  \lambda_t = -(3.05 \pm 0.75) \cdot 10^{-4}\,, \\
  \label{eq:smW}
  \Im Z_{ds}^{\rm SM} = (1.06 \pm 0.23) \cdot 10^{-4}\,, &\qquad& \Re
  Z_{ds}^{\rm SM} = -(2.42 \pm 0.59) \cdot 10^{-4}\,. 
\end{eqnarray}

The effective $\bar s d Z$ vertex contributes to the rare decays
$\kppn$, $\klpn$, $\kpe$ and $\kmm$, and also to $\epp$. Enhancing the
effective coupling $Z_{ds}$ of one order of magnitude with respect to
the SM, as suggested in ref.~\cite{CI}, corresponds to enhancing
BR($\klpn$), BR($\kppn$) and BR($\kpe$) up to two orders of magnitude
with respect to the SM. However, this enhancement also causes
BR($\kmm$) and $\epp$ to deviate strongly from the SM prediction. As
we shall see in the following, this allows us to constrain the
possible deviation of $Z_{ds}$ from its value in the SM~\cite{BS}. To
this aim, we proceed as follows: first we write down the expressions
for the above-mentioned quantities in terms of $Z_{ds}$ and of the
remaining SM contributions, assuming that the only new physics effect
in these processes comes from $Z_{ds}$. Then, in
Sect.~\ref{sec:constr}, we discuss the present constraints on $Z_{ds}$
and the maximal enhancement of rare $K$ decays compatible with these
bounds.

Our basic formulae read then as follows (see ref.~\cite{BS} for details):
\begin{eqnarray}
  {\rm BR}(\kppn) &=& 1.55 \cdot 10^{-4} \Biggl[ \biggl( \Im Z_{ds}-4 B_0
  \Im \lambda_t\biggr)^2
  + \biggl(\Re Z_{ds} + \Delta_c - 4 B_0 \Re
  \lambda_t\biggr)^2\Biggr]\,, 
  \label{eq:fkppn} \\
  \label{eq:fklpn}
  {\rm BR}(\klpn) &=& 6.78 \cdot 10^{-4} \Bigl[ \Im Z_{ds} - 4 B_0 \Im
  \lambda_t \Bigr]^2\,, \\
  \label{eq:fkpe}
  {\rm BR}(\kpe)_{\rm dir} &=& 1.19 \cdot 10^{-4} \biggl[ \Bigl( \Im
  Z_{ds} - B_0 \Im 
  \lambda_t\Bigr)^2
  + \Bigl(\Im \lambda_t + 0.08 \Im Z_{ds}\Bigr)^2\biggr]\,, \\
  \label{eq:fkmm}
  {\rm BR}(\kmm)_{\rm SD} &=& 6.32 \cdot 10^{-3} \Bigl[\Re Z_{ds} - B_0 \Re
  \lambda_t + \bar \Delta_c \Bigr]^2 \,,
\end{eqnarray}
where ${\rm BR}(\kpe)_{\rm dir}$ represents the direct CP-violating
contribution to $\kpe$, ${\rm BR}(\kmm)_{\rm SD}$ the short distance
contribution to $\kmm$, $\Delta_c = - (2.11 \pm 0.30) \cdot 10^{-4}$
and $\bar \Delta_c = - (6.54 \pm 0.60) \cdot 10^{-5}$ are the charm
contributions~\cite{nlo1} and $B_0=-0.182$ is the box diagram function
evaluated at $\mtb(m_t)=166$ GeV. Finally we isolate the $Z$
contribution to $\epp$ and proceeding as in ref.~\cite{BS} we find
\begin{eqnarray}
  \label{eq:eppz}
  \left(\frac{\varepsilon^\prime}{\varepsilon}   \right)_Z &=& \Im
  Z_{ds} \Bigl[ 1.2 - R_s \vert r_Z^{(8)}\vert B_8 ^{(3/2)}\Bigr]\,, \\
  \left(\frac{\varepsilon^\prime}{\varepsilon}   \right)_{\rm Rest}
  &=& \Im \lambda_t \biggl[-2.3 + R_s \Bigl[1.1 \vert r_Z^{(8)}\vert
  B_6^{(1/2)} + (1.0 + 0.12 \vert r_Z^{(8)}\vert )
  B_8^{(3/2)}\Bigr]\biggr]\,.\nn 
\end{eqnarray}
Here
\begin{equation}
  \label{eq:rs}
  R_s = \left[ \frac{158 {\rm MeV}}{m_s(m_c) + m_d(m_c)} \right]^2\,.
\end{equation}
$B_6^{(1/2)}$ and $B_8^{(3/2)}$ are non-perturbative parameters
describing the hadronic matrix elements of the dominant QCD-penguin
and electroweak penguin operators respectively. Finally $\vert
r_Z^{(8)}\vert$ is a calculable renormalization scheme independent
parameter in the analytic formula for $\epp$ in \cite{bratios} which
increases with $\als^{\overline {MS}}(M_Z)$ and in the range $0.116
\le \als^{\overline {MS}}(M_Z) \le 0.122$ takes the values
$  7.1 \le  \vert r_Z^{(8)}\vert \le 8.4$.
For $R_s$ we will use the range  $1 \le R_s \le 2$,
which is compatible with the most recent lattice and QCD sum rules
calculations~\cite{epp}. Similarly we will use $
  0.7 \le B_6^{(1/2)} \le 1.3$ and $0.6 \le B_8^{(3/2)} \le 1.0$
as in~\cite{epp}. 
Our treatment of $\Im \lambda_t$ and $\Re\lambda_t$ will be explained
below.

\section{Constraints on $Z_{ds}$ and results for rare K decays}
\label{sec:constr}

In deriving the bounds on $\Re Z_{ds}$ and $\Im Z_{ds}$ from $\kmm$
and $\epp$, the question arises of whether $\lambda_t$ extracted from
the standard analysis of the unitarity triangle and given in
\r{eq:smlt} is significantly modified by new physics contributions.
Following the discussion in ref.~\cite{BS}, we distinguish between two
different scenarios:
\begin{itemize}
\item[A)] New physics only manifests itself through an enhanced $\bar
  s d Z$ vertex. In this case, the standard analysis of the unitarity
  triangle is only marginally affected by new physics~\cite{BS} and we
  can use for $\lambda_t$ the values given in eq.~\r{eq:smlt}.
\item[B)] New physics affects the $\Delta F=2$ amplitudes through some
  mechanism other than $Z$-exchange. In this case the standard
  analysis of the unitarity triangle is not valid. In order to be as
  general as possible, in this scenario we only assume unitarity of
  the CKM matrix, and we obtain
  \begin{equation}
    \label{eq:ltgen}
    -1.73 \cdot 10^{-4} \le \Im\lambda_t \le 1.73 \cdot 10^{-4}\,,\qquad
    1.61 \cdot 10^{-4} \le -\Re\lambda_t \le 5.60 \cdot 10^{-4}\,.
  \end{equation}
\end{itemize}

\subsection{Results in scenario A}

The $\kmm$ branching ratio can be decomposed generally as follows:
\begin{equation}
  \label{eq:deckmm}
  {\rm BR}(\kmm)=\vert \Re A \vert^2 + \vert \Im A \vert^2\,,
\end{equation}
where $\Re A$ denotes the dispersive contribution and $\Im A$ the
absorptive one. Following the same procedure as in ref.~\cite{BS}, we
obtain $\vert \Re A_{\rm SD}\vert^2 < 2.8 \cdot 10^{-9}$.  Due to the
presence of the charm contribution, the constraint on $\Re Z_{ds}$
that one can extract from $\Re A_{\rm SD}$ depends on the sign of $\Re
Z_{ds}$. From \r{eq:fkmm} we get
\begin{equation}
  \label{eq:rewdslim}
   -5.6 \cdot 10^{-4} \le \Re Z_{ds} \le 8.0 \cdot 10^{-4}.
\end{equation}
The upper bound is obtained using $\Re\lambda_t=-3.8\cdot 10^{-4}$
and the lower one using $\Re\lambda_t=-2.3\cdot 10^{-4}$.

Having the upper bound on ${\rm BR}(\kmm)_{\rm SD}$ we can derive an
upper bound on ${\rm BR}(\kppn)$ (see ref.~\cite{BS} for details):
\begin{equation}
\label{bound}
{\rm BR}(\kppn)\le 0.229\cdot {\rm BR}(\klpn)+1.7\cdot 10^{-10}.
\end{equation}

This formula allows then to find the upper bound on ${\rm BR}(\kppn)$
once the upper bound on ${\rm BR}(\klpn)$ is known. As demonstrated in
ref.~\cite{BS}, the latter bound can be obtained from $\epp$.

The bound for $\Im Z_{ds}$ from $\epp$ depends on the sign of $\Im
Z_{ds}$. For $\Im Z_{ds}>0$ the contribution to $\epp$ is negative,
and the constraint is determined by the minimum value of $\epp$, while
the opposite happens for $\Im Z_{ds}<0$. In table \ref{tab:rare} we
report the constraints on $\Im Z_{ds}$ from $\epp$ for different
values of $(\epp)^{\rm exp}$ in the ball park of the recent
experimental results~\cite{NA31}, for $B_8^{(3/2)}=0.6~(1.0)$. These
bounds are obtained varying all the remaining parameters in
eq.~\r{eq:eppz} in the ranges given above. It turns out that the
dependence of the bounds on $B_6^{(1/2)}$ and on $R_s$ is small.
We also report in table \ref{tab:rare} the maximal values of rare $K$
decays compatible with the constraints from $\epp$ and $\kmm$. These
should be compared with the experimental data~\cite{kplus}:
 $$ {\rm BR}(\kppn) = (4.2 \,^{+9.7}_{-3.5})\cdot 10^{-10}\,,~
  {\rm BR}(\klpn) < 1.6 \cdot 10^{-6}\,, 
  ~ {\rm BR}(\kpe) < 4.3 \cdot 10^{-9} $$
and with the SM predictions
\begin{eqnarray}
  &&{\rm BR}(\klpn)=(2.8 \pm 1.1) \cdot 10^{-11}\,, \qquad
  {\rm BR}(\kppn)=(7.9 \pm 3.1) \cdot 10^{-11}\,, \nn \\
  &&{\rm BR}(\kpe)_{\rm dir}=(4.6 \pm 1.8) \cdot 10^{-12}\,. 
  \label{eq:brkpe}
\end{eqnarray}

\begin{table}[t]
  \caption{Constraints on $\Im Z_{ds}$ from $\epp$ and upper bounds on
    rare $K$ decays in scenario A for $B_8^{(3/2)}=0.6~(1.0).$}
  \label{tab:rare}
  \vspace{0.4cm}
  \begin{center}
    \begin{tabular}{|c|c|c|c|}
      \hline 
      $(\epp)^{\rm exp}_{\rm min}$ &  $1.5 \cdot 10^{-3}$ & $2 \cdot
      10^{-3}$ & $2.5 \cdot 10^{-3}$ \\\hline
      $\left(\Im Z_{ds}\right)_{\rm max}[10^{-4}]$ 
      & $2.7~(1.7)$ & $2.2~(1.4)$ & $1.6~(1.1)$ \\ \hline
      ${\rm BR}(\klpn)[10^{-11}]$ & $10.4~(5.7)$ &$7.7~(4.5)$ & 
      $5.3~(3.5)$\\ 
      ${\rm BR}(\kpe)[10^{-12}]$ & $15.0~(8.6)$ &$ 11.2~(7.1)$ &
      $8.0~(5.8)$ \\  
      ${\rm BR}(\kppn)[10^{-10}]$ & $1.9~(1.8)$ & $1.9~(1.8)$ &
      $1.8~(1.8)$ \\  \hline \hline
      $(\epp)^{\rm exp}_{\rm max}$ &  $3 \cdot 10^{-3}$ & $2.5 \cdot 10^{-3}$
      & $2 \cdot 10^{-3}$ \\ \hline 
      $\left(-\Im Z_{ds}\right)_{\rm max}[10^{-4}]$ 
      & $8.3~(4.2) $ & $6.7~(3.3)$ & $5.1~(2.5)$ \\ \hline
      ${\rm BR}(\klpn)[10^{-10}]$ & $3.9~(0.8)$ & $2.4~(0.5)$&
      $1.3~(0.2)$ \\ 
      ${\rm BR}(\kpe)[10^{-11}]$ & $7.9~(2.0)$ & $5.1~(1.3)$ &
      $2.9~(0.7)$ \\  
      ${\rm BR}(\kppn)[10^{-10}]$ & $2.6~(1.9)$ & $2.2~(1.8)$ &
      $2.0~(1.7)$ \\ \hline
    \end{tabular}
  \end{center}
\end{table}
\begin{table}[t]
  \caption{Constraints on $\Im Z_{ds}$ from $\epp$ and upper bounds on
    rare $K$ decays in scenario B for $B_8^{(3/2)}=0.6~(1.0).$}
  \label{tab:raregen}
  \vspace{0.4cm}
  \begin{center}
    \begin{tabular}{|c|c|c|c|}
      \hline
      $(\epp)^{\rm exp}_{\rm min}$ &  $1.5 \cdot 10^{-3}$ & $2 \cdot
      10^{-3}$ & $2.5 \cdot 10^{-3}$ \\ \hline 
      $\left(\Im Z_{ds}\right)_{\rm max}[10^{-4}]$ 
      & $3.0~(1.9)$ & $2.4~(1.6)$ & $1.9~(1.2)$ \\ \hline
      ${\rm BR}(\klpn)[10^{-11}]$ & $12.3~(6.7)$ &$9.3~(5.4)$ & 
      $6.7~(4.3)$\\ 
      ${\rm BR}(\kpe)[10^{-11}]$ & $1.8~(1.0)$ &$ 1.3~(0.8)$ &
      $1.0~(0.7)$ \\ 
      ${\rm BR}(\kppn)[10^{-10}]$ & $2.3~(2.2)$ & $2.2~(2.1)$ &
      $2.2~(2.1)$ \\ \hline \hline
      $(\epp)^{\rm exp}_{\rm max}$ &  $3 \cdot 10^{-3}$ & $2.5 \cdot 10^{-3}$
      & $2 \cdot 10^{-3}$ \\ \hline 
      $\left(-\Im Z_{ds}\right)_{\rm max}[10^{-4}]$ 
      & $14.9~(7.9) $ & $13.2~(7.1)$ & $11.6~(6.2)$ \\  \hline
      ${\rm BR}(\klpn)[10^{-10}]$ & $17.6~(5.7)$ & $14.2~(4.7)$&
      $11.2~(3.8)$ \\ 
      ${\rm BR}(\kpe)[10^{-10}]$ & $2.8~(0.9)$ & $2.3~(0.7)$ &
      $1.8~(0.6)$ \\
      ${\rm BR}(\kppn)[10^{-10}]$ & $6.1~(3.3)$ & $5.3~(3.2)$ &
      $4.6~(2.9)$ \\ \hline
    \end{tabular}
  \end{center}
\end{table}
\begin{figure}   
  \begin{center}
      \input{fig.tex}
    \end{center}
    \caption[]{Upper bounds on BR$(\klpn)$ in scenario B as a function
      of $\Im \lambda_t$, for negative $\Im Z_{ds}$ and for three
      different values of $(\epp)^{\rm exp}_{\rm max}$: $(\epp)^{\rm
        exp}_{\rm max} = 3 \cdot 10^{-3}$ (curve I), $(\epp)^{\rm
      exp}_{\rm max} = 2.5 \cdot 10^{-3}$ (curve II) and $(\epp)^{\rm
      exp}_{\rm max} = 2 \cdot 10^{-3}$ (curve III).}
  \label{fig:scenB}
\end{figure}

\subsection{Results in scenario B}

Using the ranges in eq.~\r{eq:ltgen} for
$\lambda_t$, we obtain the constraints on $\Im Z_{ds}$ and the
upper bounds on rare $K$ decays reported in table
\ref{tab:raregen}. Here the constraints for $\Im Z_{ds}<0$ are much
weaker than in scenario A, since now it is possible to obtain, for
$(\Im\lambda_t)_{\rm min}=-1.73\cdot 10^{-4}$, a large negative gluon
contribution which can then cancel against a large positive $Z$
contribution. Of course this situation, in which $\Im\lambda_t$ has
the opposite sign with respect to scenario A, is quite extreme, and
the upper bounds on rare $K$ decays quickly decrease as one moves from
$(\Im\lambda_t)_{\rm min}$ up, as shown in figure \ref{fig:scenB}.

\section{Summary}
\label  {sec:summary}

In this talk we have considered the possibility of an enhanced $\bar s
d Z$ vertex $Z_{ds}$. Following and updating the analysis in
ref.~\cite{BS}, we have discussed the constraints on $\Im Z_{ds}$ from
$\kmm$ and $\epp$. Our results for these bounds, together with the
upper limits on the decays $\klpn$, $\kppn$ and $\kpe$, are collected
in eq.~\r{eq:rewdslim}, tables \ref{tab:rare}-\ref{tab:raregen} and
figure \ref{fig:scenB}. Clearly the best constraints on $\Im Z_{ds}$
and $\Re Z_{ds}$ will follow in the future from precise measurements
of the theoretically clean branching ratios ${\rm BR}(\klpn)$ and
${\rm BR}(\kppn)$ respectively. Meanwhile it will be exciting to
follow the development in the improved experimental values for $\epp$,
${\rm BR}(\kmm)$, ${\rm BR}(\kppn)$, ${\rm BR}(\klpn)$ and ${\rm
  BR}(\kpe)$.

\section*{Acknowledgments}
The results presented here have been obtained in a most enjoyable
collaboration with A.J. Buras. I would like to thank the organizers
for providing such a lively atmosphere in such a beautiful
surrounding. I am also much indebted to Paolo Ciafaloni for the
patience he showed while teaching me the fundamentals of skiing and to
Andrea Donini, Stefano Rigolin and Andrea Romanino for playing billiards
late at night. This work has been supported by the German
Bundesministerium f\"ur Bildung und Forschung under contract 06 TM 874
and DFG Project Li 519/2-2.

\end{document}